\documentclass{PoS}
\usepackage{graphicx}
\usepackage{subcaption}
\usepackage{braket,amsmath,hyphenat,textcomp}
\usepackage{bm}
\usepackage{url}
\usepackage{color}
\title{First 4D lattice calculation of transport coefficient $\hat{q}$ for pure gluon plasma}

\ShortTitle{First 4D lattice calculation of transport coefficient $\hat{q}$}

\author{\speaker{Amit Kumar}\\
        Department of Physics and Astronomy, Wayne State University, Detroit, Michigan 48201, USA.\\
        E-mail: \email{kumar.amit@wayne.edu}}
      
     \author{Abhijit Majumder\\
        Department of Physics and Astronomy, Wayne State University, Detroit, Michigan 48201, USA.\\
       E-mail: \email{majumder@wayne.edu}}

\author{Chiho Nonaka\\
        Department of Physics, Nagoya University, Japan \\
       E-mail: \email{  nonaka@hken.phys.nagoya-u.ac.jp }}

\abstract{The transport coefficient $\hat{q}$ plays a pivotal role in describing  the phenomenon of jet quenching in the quark-gluon plasma (QGP) produced in  ultra-relativistic  nucleus-nucleus collisions.
It is challenging to compute this coefficient from  first principles due to its non-perturbative nature. In this article, we present an {\it ab-initio} formulation of $\hat{q}$ based on the standard techniques of  perturbative quantum chromodynamics (pQCD) and lattice gauge theory.
We construct $\hat{q}$ by considering a leading order (LO) process where a hard parton produced from the hard scattering undergoes transverse broadening due to scatterings with the thermal medium. We do an analytic continuation to the Euclidean region and use the dispersion relation to express $\hat{q}$ in terms of series of local Field-Strength-Field-Strength (FF) operators. Each term in the series is suppressed by the hard scale $q^{-}$.
Finally, we compute the local operators on the quenched SU(3) lattice and present our estimates for $\hat{q}$.
 }

\FullConference{International Conference on Hard and Electromagnetic Probes of High-Energy Nuclear Collisions\\
		30 September - 5 October 2018\\
		Aix-Les-Bains, Savoie, France}

\begin{document}
\section{Introduction}
The phenomenon of jet quenching has been widely studied as a probe of the formation of the quark-gluon plasma (QGP) in  ultra-relativistic nucleus-nucleus collisions. The leading coefficient that controls jet quenching is the jet transport coefficient $\hat{q}$ which measures the average squared transverse momentum broadening per unit length.
Previously, several attempts have been made to compute  $\hat{q}$ from phenomenology as well as first principles 
\cite{ADSCFSTcalculation1234, LTFTMajumder,
HTLcalculation,
WilsonLineApproachCalculation,   
QuenchedSU2Calculation,
MPanero2014,    
BurkeEtal, MajumderPhenomenology,
AKumarJetPuzzle,AKumarLatticeQhat}.
A hard thermal loop (HTL) perturbation theory based calculation predicts $\hat{q}$ to scale as a product of $T^3$ times  $\mathrm{log}(E/T)$ \cite{HTLcalculation}. 
A lattice gauge theory based approach has also been  developed by the authors   \cite{QuenchedSU2Calculation}  to compute $\hat{q}$ on a 4D quenched SU(2) plasma and has been extended to 4D quenched SU(3) plasma \cite{AKumarLatticeQhat}.
A phenomenology based state-of-the-art extraction of $\hat{q}$ also exists due to the work by the  JET collaboration \cite{BurkeEtal}. In this extraction,  $\hat{q}$ is assumed to scale as $T^3$.
Recently, it has been demonstrated by the authors of Ref. \cite{AKumarJetPuzzle} that $\hat{q}$ also depends on the resolution scale of the jet.
In this article, we follow the framework described in Ref. \cite{QuenchedSU2Calculation,AKumarLatticeQhat} and  present the lattice calculation of $\hat{q}$ for a 4D quenched plasma.

\section{Computing local operators on the lattice}
A framework to compute $\hat{q}$ from first principles applicable to a thermal medium was first proposed in Ref. \cite{QuenchedSU2Calculation}.  This work is further extended and applied to the quenched SU(3) plasma by the authors of Ref. \cite{AKumarLatticeQhat}. In this section, we briefly discuss the {\it{ab-initio}} formulation of $\hat{q}$. We consider a propagation of a hard quark with momenta $q=(\mu^2/2q^{-},q^{-}, 0_{\perp})\sim (\lambda^{2}Q , Q , 0)$ traveling along the negative $z$-direction, exchanging a transverse gluon [$k=(k^{+}, k^{-}, k_{\perp}) \sim (\lambda^{2} Q, \lambda^{2}Q,\lambda Q)$] with the hot plasma at temperature $T$, where $ Q >>  \Lambda_{\mathrm{QCD}}$, $\lambda << 1 $, and $\mu$ is the virtuality of the hard quark. For this leading process, one can obtain the following expression of $\hat{q}$ given as
\begin{equation}
\hat{q} = \frac{ \langle \vec{k}^{2}_{\perp} \rangle }{L}=\frac{8 \sqrt{2} \pi   \alpha_{s}(\mu^2)    }{N_c}    \int \frac{ dy^{-} d^{2} y_{\perp} } {(2\pi)^3}  d^{2} k_{\perp}  e^{ -i  \frac{\vec{k}^{2}_{\perp}}{2q^{-}} y^{-} +i\vec{k}_{\perp}.\vec{y}_{\perp} } \sum _{n}  \bra{n} \frac{e^{-\beta E_{n}}}{Z} F^{+ \perp _{\mu}}(0) F^{+}_{\perp_{\mu}}(y^{-},y_{\perp}) \ket{n},
\end{equation}
 where $N_{c}$ is the number of colors,  $\alpha_{s}(\mu^2)$ is the QCD coupling between the hard quark and the transverse gluon at a perturbative scale $\mu^2$, $L$ is the length of the box, $F^{ \mu\nu} = t^{ a} F^{ a\mu\nu}$ is the gluon field strength,   $\ket{n}$ is a state with energy $E_n$, $Z$ is the  partition function of the thermal medium, and $\beta$ is the inverse temperature. It is challenging to compute the thermal thermal expectation value of the operator $ F^{+ \perp _{\mu}}(0) F^{+}_{\perp_{\mu}}(y^{-},y_{\perp})$ due to the non-locality of the operator. However, it is possible to express $\hat{q}$ in terms of a series of local operators \cite{QuenchedSU2Calculation,AKumarLatticeQhat} :
\begin{equation}
\hat{q} = \frac{ 8\sqrt{2}\pi      \alpha_{s}(\mu^2)        }{N_c (T_{1} + T_{2})} \bra{M} F^{+ \perp_{\mu}}(0) \sum^{\infty}_{n=0}   \left(   \frac{i\sqrt{2}D_{z} }{q^{-}}\right)^{n} F^{+}_{\perp_{\mu}}(0)    \ket{M}_{(\mathrm{Thermal-Vacuum})},
\label{eq:qhatLatticeEquation}
\end{equation} 
where $T_1 +T_2 \sim 2T-4T$, and  $D_z$ is the covariant derivative in the $z$-direction.
The above expression of the transport coefficient $\hat{q}$ (Eq. \ref{eq:qhatLatticeEquation}) is valid for both pure gluon and quark-gluon plasma.  It is interesting to highlight that a similar kind of operator product expansion containing covariant derivative $D_{z}$ has been found by the author of Ref. \cite{XJiPartonPDF} in  the analysis of the parton distribution function in the Euclidean space.
In our first attempt, we have computed $\hat{q}$ by considering the first two terms in the series (Eq.  \ref{eq:qhatLatticeEquation}) and have ignored higher order terms. In the Euclidean space these operators are
$
F^{+ \perp_{\mu}}(0) F^{+}_{ \perp_{\mu}}(0)  \overset{}{\longrightarrow} \frac{1}{2}  \left[    \sum^{2}_{i=1}( F^{3i}F^{3i} - F^{4i}F^{4i})    + i \sum^{2}_{i=1} (F^{4i}F^{3i} + F^{3i}F^{4i})   \right]
$
and 
$
 F^{+ \perp_{\mu}} \left(  \frac{i\sqrt{2}D_{z}}{q^{-}}  \right)  F^{+}_{ \perp_{\mu}}  
  \overset{}{\longrightarrow}   \frac{ \sqrt{2}}{2q^-}   \left[  i  \sum^{2}_{i=1}( F^{3i} D_{z} F^{3i} - F^{4i} D_{z} F^{4i})    -   \sum^{2}_{i=1} (F^{4i} D_{z} F^{3i} + F^{3i} D_{z}  F^{4i})  \right].
$
\begin{figure}[h]
  \centering
  \begin{subfigure}[b]{0.49\linewidth}
    \includegraphics[width=\linewidth]{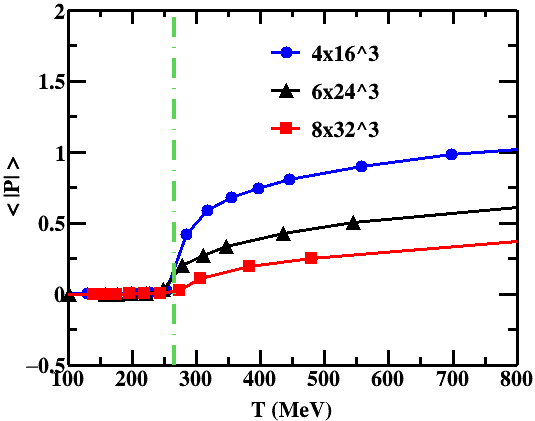}
    \caption{}
  \end{subfigure}
  \begin{subfigure}[b]{0.49\linewidth}
    \includegraphics[width=\linewidth]{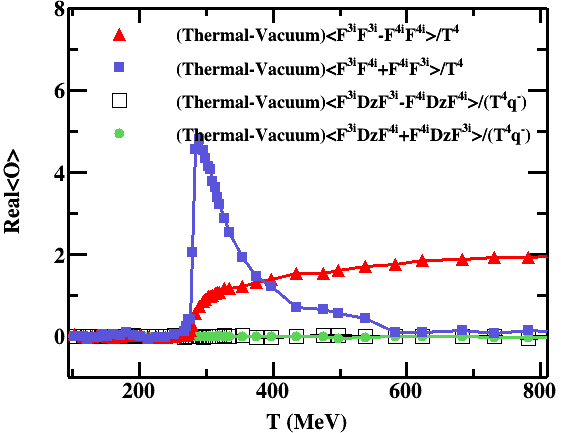}
    \caption{}
  \end{subfigure}
  
   \begin{subfigure}[b]{0.49\linewidth}
    \includegraphics[width=\linewidth]{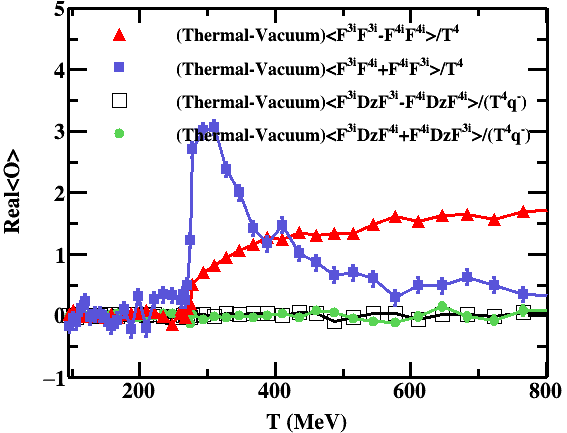}
    \caption{}
  \end{subfigure}
  \begin{subfigure}[b]{0.49\linewidth}
    \includegraphics[width=\linewidth]{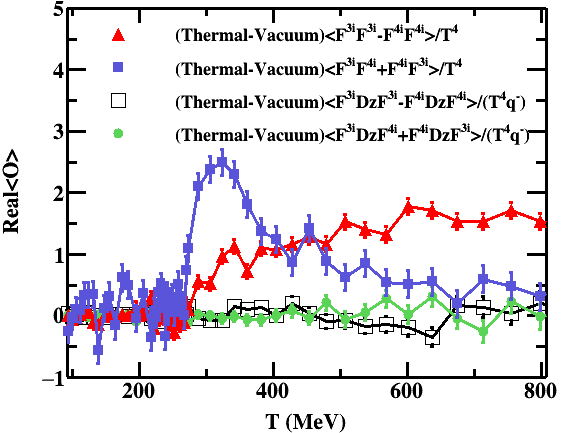}
    \caption{}
  \end{subfigure}
  \caption{Average value of operators as a function of temperature for pure gluon plasma. We used $q^{-}$ as 20 GeV. (a) The Polyakov loop for $n_{\tau}=$ 4, 6, and 8.  (b)  The real part of FF correlators for $n_{\tau}=4$ and $n_{s}=16$. (c) The real part of FF correlators for $n_{\tau}=6$ and $n_{s}=24$. (d) The real part of FF correlators for $n_{\tau}=8$ and $n_{s}=32$.}
  \label{fig:OperatorsWrtTemperature}
\end{figure}
We have evaluated the operators for the quenched case using the Wilson gauge action and left the unquenched calculation for future work. 
In our Monte Carlo (MC) calculations, we have taken statistical average over 5000 gauge configurations. We have used the standard  heat-bath algorithm to generate the gauge configurations. 
We have computed operators for $n_{\tau}=$ 4, 6, 8 with $n_{s}=4n_{\tau}$ 
as a function of the temperature.  To do this, we set the scale on the lattice using a non-perturbative renormalization group (RG) equation, that utilizes the two loop RG equation  \cite{SU3ScaleSettings}, given as $
a_{L} = \frac{f}{\Lambda_{L}} \left[ \frac{11g^2}{16\pi^2} \right]^{\frac{-51}{121}}  \exp  \left[ \frac{-8\pi^2}{11g^2} \right],
$  
 where $g$ is the bare lattice coupling, $a_{L}$ is the lattice spacing, $f$ is non-perturbative correction, and $\Lambda_{L}$ is a dimensionful parameter on the lattice.  This determines the
temperature  $T$, given by $1/(n_{\tau} a_{L})$. 
To estimate the factor $f$, we compute  Polyakov loop ($\langle  |P| \rangle $) and adjust the parameter $f$ such that $T_{c}/\Lambda_{L}$ does not depend on the  bare lattice coupling  $g$. Here, $T_{c}\sim 265 $ MeV is the critical temperature for pure SU(3) gauge theory \cite{SU3ScaleSettings}. 
The Polyakov loop ($P$)  is defined  as
$
P = \frac{1}{n_{x}n_{y}n_{z}}tr \left[  \sum_{ \vec{r}} \prod ^{n_{\tau}-1}_{n=0} U_{4}(na,\vec{r})   \right],
$
where $U_{4}(na,\vec{r})$ is a gauge link in the temporal ($\tau$) direction. 

We present the average value of the Polyakov loop in  Fig. \ref{fig:OperatorsWrtTemperature}(a) after performing the non-perturbative renormalization. The green vertical line indicates the critical temperature for pure SU(3) gauge theory [see Fig. \ref{fig:OperatorsWrtTemperature}(a)].
We also show the expectation value of the FF correlators as a function of temperature in Figs. \ref{fig:OperatorsWrtTemperature}(b), \ref{fig:OperatorsWrtTemperature}(c) and \ref{fig:OperatorsWrtTemperature}(d) for $n_{\tau}=$ 4, 6 and 8, respectively. The FF correlators are scaled by $T^{4}$ or $T^4q^{-}$ to make them dimensionless. The uncrossed correlator  $\langle   F^{3i}F^{3i}-F^{4i}F^{4i} \rangle $ (red curve) exhibits a rapid transition near temperature $T\in (250,350)$ MeV and is dominant at high temperatures compared to rest of the FF correlators [see Figs. \ref{fig:OperatorsWrtTemperature}(b,c,d)]. 
However, the crossed correlator   $\langle  F^{3i}F^{4i}+F^{4i}F^{3i} \rangle $ (blue curve) exhibits a peak-like behavior near the transition region $T \in (250,350)$ MeV and is suppressed at high temperatures [see Figs. \ref{fig:OperatorsWrtTemperature}(b,c,d)]. We also observe that both the FF correlators  containing covariant derivatives (green and black curves) are suppressed at all temperatures [see Figs. \ref{fig:OperatorsWrtTemperature}(b,c,d)]. 

\section{Results and discussions}
\begin{figure}[h!]
  \centering
  \begin{subfigure}[b]{0.45\linewidth}
    \includegraphics[width=\linewidth]{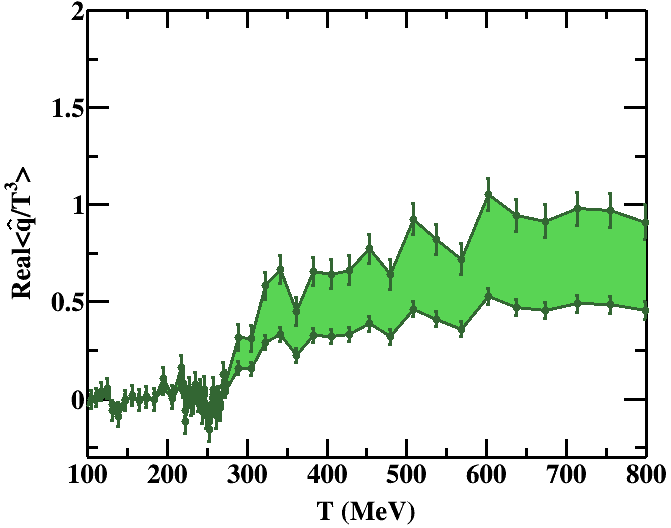}
    \caption{}
  \end{subfigure}
  \begin{subfigure}[b]{0.45\linewidth}
    \includegraphics[width=\linewidth]{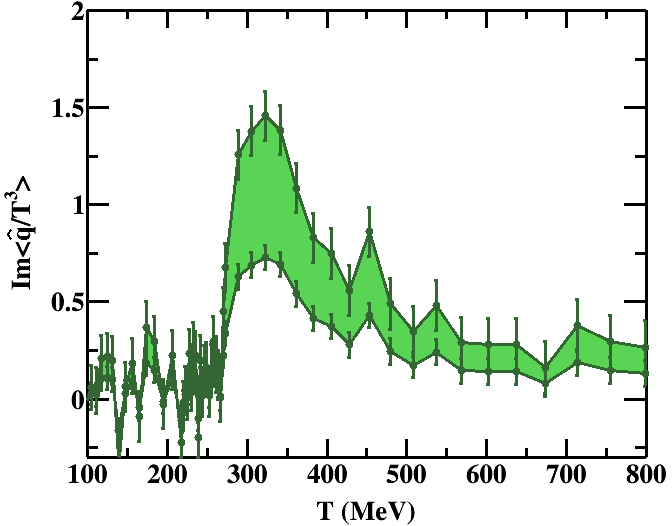}
    \caption{}
  \end{subfigure}
  \caption{Temperature dependence of $\hat{q}/T^3$ for hard quark propagating through a pure gluon plasma. (a) The real part of $\hat{q}/T^3$. (b) The Imaginary part of $\hat{q}/T^3$.}
  \label{fig:qhatT3WrtT}
\end{figure}

 To compute $\hat{q}$ [ Eq. \ref{eq:qhatLatticeEquation}],  we have taken the QCD coupling  constant  $\alpha_{s}=0.2$  $(\mu^2 \simeq 8$ $\mathrm{ GeV}^2)$ and the hard quark energy $q^{-}=20$ GeV.  Since, there is only one energy, or rather virtuality associated with the hard quark in the calculation, one value of $(q^{-},\mu^2)$ have been used. We present our final result of $\hat{q}$ in Fig. \ref{fig:qhatT3WrtT}   computed on the quenched SU(3) lattice. 
The figure \ref{fig:qhatT3WrtT}(a) represents the real part of $\hat{q}/T^3$ which exhibits a transition region around $T\in (250,350)$ MeV. At high temperature, our lattice calculation constrains the extracted value of $\hat{q}/T^3$ to be $\sim 0.5-1$.
Naively, one expects the imaginary part of $\hat{q}/T^3$ to be zero at all temperatures, but our lattice calculations show a peak-like behavior near the transition region $T\in (250,350)$ MeV, and it goes to zero at high temperatures [see Fig. \ref{fig:qhatT3WrtT}(b)].
 The vacuum subtracted expectation value of crossed correlator $\langle  F^{3i}F^{4i}+F^{4i}F^{3i}  \rangle $ is non-zero near the transition region. This gives rise the imaginary part to  $\hat{q}$.
 Note, the operator $\langle  F^{3i}F^{4i}+F^{4i}F^{3i}  \rangle $ represents
$(\vec{E} \times \vec{B})_{z} $, a Poynting vector in the $z$-direction 
which causes the damping effects for the thermal gluons originating from the plasma.

\section{Summary and Outlook}   
In this article, we have  established an {\it ab-initio} framework of $\hat{q}$ applicable to 4D hot quark-gluon plasma and extended the work presented in  Ref. \cite{QuenchedSU2Calculation,AKumarLatticeQhat}. 
 We computed FF correlators on a quenched SU(3) lattice for different lattice sizes and demonstrated the scaling behavior. Our quenched lattice calculation constrains $\hat{q}/T^3 \sim 0.5-1$ at high temperature.

In the next step, we will compute the FF correlators on the unquenched SU(3) lattice and extend this study to the full quark-gluon plasma.
We will also include the corrections to $\hat{q}$ from the medium-induced radiative splitting.  We will also extend the study to improved actions.
\\
{\bf Acknowledgment}
This work was supported in part by the National Science Foundation under the grant No. ACI-1550300 within the framework of the JETSCAPE collaboration, and in part by US department of energy, office of science, office of nuclear physics under grant No. DE-SC0013460.

\end{document}